\documentclass[twocolumn,english,secnumarabic,amssymb,nobibnotes,aps,pre,showpacs]{revtex4-1}
\usepackage[T1]{fontenc}
\usepackage[latin9]{inputenc}
\usepackage{geometry}
\usepackage{amsmath}
\usepackage{amssymb}

\makeatletter

\providecommand{\tabularnewline}{\\}

\@ifundefined{textcolor}{}
{%
 \definecolor{BLACK}{gray}{0}
 \definecolor{WHITE}{gray}{1}
 \definecolor{RED}{rgb}{1,0,0}
 \definecolor{GREEN}{rgb}{0,1,0}
 \definecolor{BLUE}{rgb}{0,0,1}
 \definecolor{CYAN}{cmyk}{1,0,0,0}
 \definecolor{MAGENTA}{cmyk}{0,1,0,0}
 \definecolor{YELLOW}{cmyk}{0,0,1,0}
 }

\usepackage{amsfonts}
\usepackage{epsfig}

\usepackage{psfrag}

\usepackage{colordvi}
\usepackage{color}

\newcommand{\sfrac}[2]{\left[\begin{smallmatrix} #1\\ #2 \end{smallmatrix}\right]}

\makeatother

\usepackage{babel}
\begin{document}

\title{Equivalence between non-bilinear spin-$S$ Ising model and Wajnflasz
model}

\author{Onofre Rojas\footnote{email:ors@dex.ufla.br; Phone: +5535 38291954, Fax: +5535 3829-1961.}
and S. M. de Souza}

\affiliation{Departamento de Ciências Exatas, Universidade Federal de Lavras.
CP 3037, 37200-000, Lavras, MG, Brazil.}
\begin{abstract}
We propose the mapping of polynomial of degree $2S$ constructed as
a linear combination of powers of spin-$S$ (for simplicity, we called
as spin-$S$ polynomial) onto spin-crossover state. The spin-$S$
polynomial in general can be projected onto non-symmetric degenerated
spin up (high-spin) and spin down (low-spin) momenta. The total number
of mapping for each general spin-$S$ is given by $2(2^{2S}-1)$.
As an application of this mapping, we consider a general non-bilinear
spin-$S$ Ising model which can be transformed onto spin-crossover
described by Wajnflasz model. Using a further transformation we obtain
the partition function of the effective spin-1/2 Ising model, making
a suitable mapping the non-symmetric contribution leads us to a spin-1/2
Ising model with a fixed external magnetic field, which in general
cannot be solved exactly. However, for a particular case of non-bilinear
spin-$S$ Ising model could become equivalent to an exactly solvable
Ising model. The transformed Ising model exhibits a residual entropy,
then it should be understood also as a frustrated spin model, due
to competing parameters coupling of the non-bilinear spin-$S$ Ising
model.
\end{abstract}



\maketitle

\section{Introduction}

One of the topics of great interest, in statistical physics and mathematical
physics are the exact solvable models. This is the case, i.e. for
spin-1/2 Ising model without magnetic field solved at first in 1944
by Onsager\cite{onsager}, since that, the Ising model was widely
investigated using several approaches. On the other hand, higher order
spin or even spin-1/2 Ising model with external magnetic field are
challenging issues in nowadays. Further exact solution were obtained
only in a very limited cases, mainly the honeycomb lattices\cite{Horiguchi-Wu,Kolesik}.
Some exact results has been obtained with restricted parameter, investigated
by Mi and Yang\cite{MiYang} using a non-one-to-one transformation\cite{Kolesik}.
Therefore the non-bilinear spin-$S$Ising model that satisfy this
transformation should exhibit frustrated states.

The half-odd-integer spin Ising model already has been discussed previously
by Tang\cite{tang}. Using the method proposed by Wu\cite{wu}, Izmailian
\cite{izmailian} obtained an exact solution for a spin-3/2 lattice
on square lattice with only nearest interaction or two body interaction
spin, Izmailian and Ananikian\cite{izm-anani} also has been obtained
an exact solution for a honeycomb lattice with spin-3/2. A particular
solution of these models could be obtained using the method proposed
by Joseph\cite{joseph} where any spin-$S$ could be projected onto
a spin-1/2 Ising model. Another interesting method to map the spin-$S$
model onto spin-1/2 Ising model has been proposed by Horiguchi\cite{horiguchi}.
More recently we have obtained a set of rigorous mapping for half-odd-integer
spin onto spin-1/2\cite{lett-exact}, where we have used a direct
mapping, half of spin momenta are projected onto spin down, while
the remaining half spin momenta are projected onto spin up, we will
call this process as spin-$S$ projection with symmetric degeneracy,
this mapping is a non-one-to-one mapping.

However, when we consider an integer spin we cannot perform the mapping
symmetrically, this issue will be considered in this letter. Then
using a non-symmetric projection we will discuss the mapping for a
general spin-$S$ polynomial onto a spin-crossover state. 

On the other hand, the spin crossover (SC), sometimes called as spin
transition, is a phenomenon that occurs in some metal (i.e. Fe and
Co) complexes wherein the spin state of the complex changes due to
external perturbation such as a variation of temperature, pressure,
light irradiation or an influence of a magnetic field\cite{Real}.
The spin states of the atoms can change between the high-spin (HS)
state and low-spin (LS) state as a result of external stimuli, which
can be understand as a non-symmetric degeneracy between HS and LS
states. In this sense we find a equivalence between spin-$S$ polynomial
and spin-crossover state. Another interesting equivalence could be
also to that the metastable structure of a charge transfer phase transition,
discussed by Miyashita et al.\cite{miyashita}, where static metastability
exist in a study of the charge transfer transition in the material
$\mathrm{(nC_{3}H_{7})_{4}N[Fe^{II}Fe^{III}(dto)_{3}]}$ (dto=$\mathrm{C_{2}O_{2}S_{2}}$).

The outline of this report is as follow: In sec. 2 we present the
mapping of spin-$S$ polynomial onto spin-crossover state, in sec.
3 we apply to a a non-bilinear spin-$S$ Ising model and its relation
to the metastable state, while in sect 4 we discuss the exactly solvable
case. Finally in sec. 5 we present our conclusions.

\section{The spin-$S$ polynomial transformation onto spin-crossover state}

In order to show the equivalence between spin-$S$ polynomial and
spin-crossover state. Let us start considering as an example the projection
of spin-3/2 polynomial, as follow, 
\begin{align}
\sigma_{m}^{(\frac{3}{2})}(s) & =\alpha_{0,m}+\alpha_{1,m}s+\alpha_{2,m}s^{2}+\alpha_{3,m}s^{3},\label{eq:sp3/2}
\end{align}
where $\alpha_{i,m}$ with $i=0,\dots,3$, are the coefficients to
be determined using the projection of spin-3/2 onto spin-1/2, whereas
by $m$ we mean the number of solutions or projections$1\leqslant m\leqslant2(2^{3}-1)$.
The polynomial spin-3/2 of eq.\eqref{eq:sp3/2} can be rewritten alternatively
by

\begin{equation}
\boldsymbol{\sigma}_{m}^{(\frac{3}{2})}=\mathsf{s}^{(\frac{3}{2})}\mathsf{\boldsymbol{\alpha}}_{m},
\end{equation}
where 
\begin{equation}
\mathsf{s}^{(\frac{3}{2})}=\left(\begin{array}{cccc}
1 & s & s^{2} & s^{3}\end{array}\right),\enskip\boldsymbol{\alpha}_{m}=\left(\begin{array}{c}
\alpha_{0,m}\\
\alpha_{1,m}\\
\alpha_{2,m}\\
\alpha_{3,m}
\end{array}\right)\mathsf{.}
\end{equation}
 To find the coefficients of the polynomial spin-3/2 eq.\eqref{eq:sp3/2},
we use the following equation,

\begin{equation}
\mathsf{\boldsymbol{\alpha}}_{m}=\mathsf{V}^{-1}\mathsf{P}_{m}.
\end{equation}
 where

\begin{equation}
\mathsf{V}=\left(\begin{array}{cccc}
1 & -\frac{3}{2} & \frac{9}{4} & -\frac{27}{8}\\
1 & -\frac{1}{2} & \frac{1}{4} & -\frac{1}{8}\\
1 & \frac{1}{2} & \frac{1}{4} & \frac{1}{8}\\
1 & \frac{3}{2} & \frac{9}{4} & \frac{27}{8}
\end{array}\right),\enskip\mathsf{P}_{m}=\left(\begin{array}{c}
P_{0,m}\\
P_{1,m}\\
P_{2,m}\\
P_{3,m}
\end{array}\right),
\end{equation}
each vector $\mathsf{P}_{m}$ are defined as a column vector of matrix
$\mathsf{P}$,

\begin{equation}
\mathsf{P}=\left(\begin{array}{ccccc}
-1 & -1 & 1 & 1 & 1\\
-1 & 1 & -1 & 1 & 1\\
1 & -1 & -1 & 1 & -1\\
1 & 1 & 1 & -1 & 1
\end{array}\right),
\end{equation}
the elements of matrix $\mathsf{P}$, means the projection spin-3/2
polynomial onto spin up (+1) or high-spin (HS) and spin down (-1)
or low-spin (LS). Each vector $\mathsf{P}_{m}$ (column of matrix
$\mathsf{P}$) are the permutations of at least one spin up and one
spin down, here the columns of matrix $\mathsf{P}$ only represents
the non-equivalent configurations of spin projections, for this case
we have 5 \textquotedbl{}representative\textquotedbl{} configurations.
We should recover easily the remaining configuration taken into account
the exchange of spin $s$, $\sigma_{m}^{(\frac{3}{2})}(s)\leftrightarrow\sigma_{m}^{(\frac{3}{2})}(-s)$
and the global inversion of the polynomial $\sigma_{m}^{(\frac{3}{2})}(s)\leftrightarrow-\sigma_{m}^{(\frac{3}{2})}(s)$.
Therefore the spin-3/2 polynomial mapping onto spin-1/2 are expressed
by

\begin{align}
\sigma_{1}^{(\frac{3}{2})}(s) & =\tfrac{13}{6}s-\tfrac{2}{3}s^{3},\label{eq:sol321}\\
\sigma_{2}^{(\frac{3}{2})}(s) & =-\tfrac{7}{3}s+\tfrac{4}{3}s^{3},\label{eq:sol322}\\
\sigma_{3}^{(\frac{3}{2})}(s) & =-\tfrac{5}{4}+s^{2},\label{eq:4}\\
\sigma_{4}^{(\frac{3}{2})}(s) & =\tfrac{9}{8}+\tfrac{1}{12}s-\tfrac{1}{2}s^{2}-\tfrac{1}{3}s^{3},\\
\sigma_{5}^{(\frac{3}{2})}(s) & =-\tfrac{1}{8}-\tfrac{9}{4}s+\tfrac{1}{2}s^{2}+s^{3}.\label{eq:sol325}
\end{align}

The mappings given in \eqref{eq:sol321}-\eqref{eq:4} already were
considered in reference\cite{izmailian,lett-exact}, which corresponds
to symmetric degeneracy mapping (the first 3 column of vector $\mathsf{P}$).
The remaining solutions corresponds to non-symmetric degeneracy, i.e.
three spin momenta are projected onto -1 (LS), whereas the remaining
spin moment is projected onto +1 (HS), or vice-verse. These solutions
have not been considered yet in the literature. 

The next transformation that we discuss could be the spin-2 polynomial
onto spin-1/2. It is not possible to map onto a spin-1/2 with symmetric
degeneracy\cite{lett-exact} because, we have five magnetic momenta
to be mapped onto two eigenvalues $\pm1$. Then the only possibility
is to map by means of non-symmetric projection onto spin-1/2, this
kind of mapping lead us to 30 polynomials. However we only need to
obtain 9 \textquotedbl{}representative\textquotedbl{} polynomials,
which are tabulated in table 1. Once again using the exchange of magnetization
$\sigma_{m}^{(2)}(s)\leftrightarrow\sigma_{m}^{(2)}(-s)$ and the
global inversion of the polynomial $\sigma_{m}^{(2)}(s)\leftrightarrow-\sigma_{m}^{(2)}(s)$,
we could obtain easily the remaining projections. 

\begin{widetext}

\begin{table}
\begin{tabular}{|l|c|c|c|c|}
\hline 
Spin-2 polynomial  & Proj. to (+1)  & Proj. to (-1) & $g(+1)$  & $g(-1)$\tabularnewline
\hline 
\hline 
$\sigma_{1}^{(2)}(s)=-\tfrac{7}{6}s-\tfrac{5}{4}s^{2}+\tfrac{1}{6}s^{3}+\tfrac{1}{4}s^{4}+1$ & $-2,-1,0$  & $1,2$ & 3  & 2 \tabularnewline
\hline 
$\sigma_{2}^{(2)}(s)=-\tfrac{4}{3}s+\tfrac{7}{6}s^{2}+\tfrac{1}{3}s^{3}-\tfrac{1}{6}s^{4}-1$  & $2,-1,-2$  & $0,1$  & 3  & 2 \tabularnewline
\hline 
$\sigma_{3}^{(2)}(s)=\tfrac{1}{6}s^{2}-\tfrac{1}{6}s^{4}+1$  & $1,0,-1$  & $2,-2$  & 3  & 2\tabularnewline
\hline 
$\sigma_{4}^{(2)}(s)=\tfrac{1}{6}s+\tfrac{31}{12}s^{2}-\tfrac{1}{6}s^{3}-\frac{7}{12}\, s^{4}-1$  & $-2,1,-1$  & $0,2$  & 3 & 2\tabularnewline
\hline 
$\sigma_{5}^{(2)}(s)=-\tfrac{8}{3}s^{2}+\tfrac{2}{3}s^{4}+1$  & $2,0,-2$  & $1,-1$  & \multicolumn{1}{c|}{3} & 2\tabularnewline
\hline 
$\sigma_{6}^{(2)}(s)=\tfrac{3}{2}s-\tfrac{5}{4}s^{2}-\tfrac{1}{2}s^{3}+\tfrac{1}{4}s^{4}+1$  & $-2,0,1$  & $-1,2$  & 3 & 2\tabularnewline
\hline 
$\sigma_{7}^{(2)}(s)=\tfrac{1}{6}s+\tfrac{1}{12}s^{2}-\tfrac{1}{6}s^{3}-\tfrac{1}{12}s^{4}+1$  & $-2,1,0,-1$  & $2$  & 4 & 1\tabularnewline
\hline 
$\sigma_{8}^{(2)}(s)=-\tfrac{4}{3}s-\tfrac{4}{3}s^{2}+\tfrac{1}{3}s^{3}+\tfrac{1}{3}s^{4}+1$  & $2,-1,0,-2$  & $1$  & 4 & 1\tabularnewline
\hline 
$\sigma_{9}^{(2)}(s)=\tfrac{5}{2}s^{2}-\tfrac{1}{2}s^{4}-1$  & $2,1,-1,-2$  & 0 & 4 & 1\tabularnewline
\hline 
\end{tabular}

\caption{The projection spin-2 onto a non-symmetric spin-1/2 or spin-crossover
state. By $g(+1)$ we mean the degeneracy of the spin up (HS), whereas
$g(-1)$ means the degeneracy of spin down (LS).}
\end{table}

\end{widetext}

In general the projection of spin-$S$ polynomial onto $\sigma(s)$
with non-symmetric spin degeneracy, we assume the following spin-$S$
polynomial,
\begin{align}
\sigma^{(S)}(s)=\sum_{j=0}^{2S}\alpha_{j}^{(S)}s^{j},
\end{align}
 where the coefficients $\alpha_{j}^{(S)}$ of the polynomial will
be determined after projecting onto spin-1/2.

To perform the spin-$S$ polynomial projection we consider the Vandermonde
matrix $\mathsf{V}^{(S)}$ with \textit{equidistant nodes} $[-S,S]$,
whose elements of the node are $x_{j}$ which corresponds just to
the magnetic momenta of the spin-$S$, the elements of the matrix
could be expressed appropriately as $x_{j}=-S+j$, with $j=0,1,2,\dots,2S$,
the explicit representation of the Vandermonde matrix is given by,
\begin{align}
\mathsf{V}^{(S)}=\begin{pmatrix}1 & x_{0} & x_{0}^{2} & x_{0}^{3} & \dots & x_{0}^{2S}\\
1 & x_{1} & x_{1}^{2} & x_{1}^{3} & \dots & x_{1}^{2S}\\
\vdots & \vdots & \vdots & \vdots & \ddots & \vdots\\
1 & x_{2S-1} & x_{2S-1}^{2} & x_{2S-1}^{3} & \dots & x_{2S-1}^{2S}\\
1 & x_{2S} & x_{2S}^{2} & x_{2S}^{3} & \dots & x_{2S}^{2S}
\end{pmatrix},\label{vandermonde}
\end{align}
and 

\begin{align}
\mathsf{\boldsymbol{\alpha}}_{m}^{(S)}=\begin{pmatrix}\alpha_{0,m}\\
\alpha_{1,m}\\
\vdots\\
\alpha_{2S-1,m}\\
\alpha_{2S,m}
\end{pmatrix},\quad\mathsf{P}_{m}^{(S)}=\begin{pmatrix}P_{0,m}\\
P_{1,m}\\
\vdots\\
P_{2S-1,m}\\
P_{2S,m}
\end{pmatrix},\label{vandermonde-1}
\end{align}
we also define the vector $\mathsf{\boldsymbol{\alpha}}_{m}^{(S)}$
to represent the coefficients for the all possible spin-$S$ polynomial,
while the elements of the vector $\mathsf{P}^{(S)}$ represents the
projection with non-symmetric degeneracy HS and LS, which can be expressed
by

\begin{align}
\left(\mathsf{P}_{m}^{(S)}\right)^{T}=\mathcal{P}\big(\underbrace{1,\dots,1}_{r\text{ times}},\underbrace{-1,\dots,-1}_{2S+1-r\text{ times}}\big),\label{mP}
\end{align}
by $\mathcal{P}$ we mean any permutation of the elements of $\mathsf{P}^{(S)}$,
with $r$ projections onto spin up (HS) and $2S+1-r$ projections
onto spin down (LS), assuming $r=\{1,\dots,2S\}$. For a given $r$
we have $\binom{2S}{r}$ permutations, which projects the spin-$S$
polynomial onto spin-1/2. We can verify that all matrix defined above
have $2S+1$ dimension.

In order to project the spin-$S$ polynomial onto $\pm1$ values with
non-symmetric degeneracy, we can use the matrix notation, so, the
following algebraic system equation becomes, 
\begin{align}
\mathsf{P}_{m}^{(S)}=\mathsf{V}^{(S)}\mathsf{\boldsymbol{\alpha}}_{m}^{(S)}.
\end{align}

The number of projection of spin-$S$ polynomials that we can obtain
are given by the permutations of the elements of the vector \eqref{mP},
for each $r=\{1,\dots,2S\}$. Therefore, the total number of solutions
for each spin-$S$ is given by $\binom{2S+1}{1}+\binom{2S+1}{2}+\dots+\binom{2S+1}{2S}=2(2^{2S}-1)$,
thus $m=\{1,\ldots,2(2^{2S}-1)\}$. 

Using the matrix notation, we are able to write in general the spin-$S$
polynomial, 
\begin{align}
\boldsymbol{\sigma}_{m}^{(S)}=\mathsf{s}^{(S)}\mathsf{\boldsymbol{\alpha}}_{m}^{(S)}=\mathsf{s}^{(S)}\left(\mathsf{V}^{(S)}\right)^{-1}\mathsf{P}_{m}^{(S)}.
\end{align}

The inverse of the matrix $\mathsf{V}^{(S)}$ could be solved using
the recursive equation presented recently by Eisinberg \textit{et
al.}\cite{eisinberg}, where was discussed a generic algorithm to
obtain the elements of the inverse of Vandermonde matrix $\mathsf{V}^{(S)}$.
Therefore the elements of the matrix $\mathsf{V}^{(S)}$ are rewritten
conveniently as in reference \cite{phys-A-09}, which reads as 
\begin{align}
\widetilde{v}_{i,j}^{(S)}= & \tfrac{(-1)^{i+j}}{(2S+1-j)!(j-1)!}\sum_{k=1}^{2S+1}(-S-1)^{k-i}\binom{k}{i}\times\nonumber \\
 & \left|\sfrac{2S+2}{k+1}\right|F_{i+1}^{1,i-k}\left(1-\tfrac{j}{S+1}\right),\label{invs-elem}
\end{align}
where $\sfrac{.}{.}$ represent the first kind of Stirling number,
whereas ${\rm F}_{i+1}^{1,i-k}$ represents the hyper-geometric function\cite{abramowitz}.

Using the elements of inverse matrix $\widetilde{v}_{i,j}^{(S)}$,
we are able to write the coefficient for each polynomial $\sigma(s)$,
\begin{align}
\alpha_{i,m}^{(S)} & =\sum_{j=0}^{2S}\widetilde{v}_{i,j}^{(S)}P_{j,m}^{(S)},
\end{align}
note that here $\alpha_{i,m}^{(S)}$ are the elements of vector $\mathsf{\boldsymbol{\alpha}}_{m}^{(S)}$. 

The non-symmetric degeneracy projection of spin-$S$ polynomial onto
a spin-1/2 is given by 

\begin{equation}
g_{r}(\sigma_{m})=\begin{cases}
r; & \sigma_{m}=-1\\
2S+1-r;\quad & \sigma_{m}=1
\end{cases}.
\end{equation}

When the spin-$S$ is half-odd-integer, we could recover the solution
already obtained in reference \cite{lett-exact} as a particular case
of our result, when the degeneracy becomes symmetric.

\section{The non-bilinear spin-$S$ Ising model mapping onto Wajnflasz model}

As an application of this mapping, let us consider the non-bilinear
spin-$S$ Ising model with two-body and high-order interaction term,
whose Hamiltonian for arbitrary spin-$S$, can be written as

\begin{equation}
\mathcal{H}_{S}=\sum_{<i,j>}\sum_{k_{1}=1}^{2S}\sum_{k_{2}=1}^{2S}K_{k_{1},k_{2}}s_{i}^{k_{1}}s_{j}^{k_{2}}-\sum_{i}\sum_{k=1}^{2S}B_{k}s_{i}^{k},\label{eq:Ham-sp-S}
\end{equation}
with $K_{k_{1},k_{2}}$ being the non-bilinear interaction terms,
while by $B_{k}$ corresponds to the high order anisotropy coupling.
The $<i,j>$ means the summation over the pairs of nearest-neighbor
sites.

In order to discuss the equivalence between non-bilinear spin-$S$
Ising model and the Wajnflasz model\cite{wanjflasz}, let us describe
the Ising-like model or Wajnflasz model\cite{wanjflasz}.

\subsection{The Ising-like model}

In order to describe the spin-crossover transition, we can use the
Wajnflasz model\cite{wanjflasz,varret,boujheddaden}, where this model
take into account the HS state and LS state, modeled simply by a nearest-neighbor
interaction between sites.

Hence the Hamiltonian with arbitrary spin-$S$ Ising model can be
mapped onto an effective spin-1/2 Ising-like model, which read as
\begin{align}
\mathcal{H}_{m}^{\prime}(\{\sigma\})=\sum_{<i,j>}J\sigma_{m}(s_{i})\sigma_{m}(s_{j})-h\sum_{i}\sigma_{m}(s_{i}),\label{hamt-s}
\end{align}
 where $J$ is the effective spin-crossover interaction parameter,
and whereas $h$ corresponds to the effective external magnetic field
or the energy difference between HS and LS states.

With aim of the eqs. \eqref{eq:Ham-sp-S} and \eqref{hamt-s} becomes
equivalent, we need to impose the following condition $\mathcal{H}_{S}=\mathcal{H}_{m}^{\prime}(\{\sigma\})-\mathcal{E}_{0}^{\prime}$.
Where the parameters must satisfy the relation below 

\begin{align}
\mathcal{E}_{0}^{\prime}= & MJ\alpha_{0,m}^{2}-Nh\alpha_{0,m},\\
B_{k}= & (h-\gamma J\alpha_{0,m})\alpha_{k,m},\quad k\geqslant1,\\
K_{k_{1},k_{2}}= & J\alpha_{k_{1},m}\alpha_{k_{2},m},\quad k_{1}\geqslant1\quad\text{{and}}\quad k_{2}\geqslant1,
\end{align}
 with $M$ being the total number of nearest-neighbor spin pairs and
$N$ being the total number of sites, while $\gamma$ corresponds
to the coordination number of the lattice.

To study thermodynamics properties, we have to compute the partition
function of Ising-like model, 
\begin{align}
\mathcal{Z}(\beta)=\sum_{\{s_{i}\}}\exp(-\beta\mathcal{H}_{S}),\label{Zdef}
\end{align}
where $\beta=1/k_{B}T$, with $k_{B}$ being the Boltzmann constant
and $T$ the absolute temperature. 

Similar to that was discussed by Mi and Yang\cite{MiYang}, the eq.\eqref{Zdef}
can be rewritten as follow, 
\begin{align}
\mathcal{Z}_{m}(\beta)= & \mathrm{e}^{\beta\mathcal{E}_{0}^{\prime}}\sum_{\{\sigma_{i}\}=\pm1}g_{r}(\sigma_{m,1})g_{r}(\sigma_{m,2})\dots g_{r}(\sigma_{m,N})\times\nonumber \\
 & \exp(-\beta\mathcal{H}_{m}^{\prime}(\{\sigma\}))\\
= & \mathrm{e}^{\beta\mathcal{E}_{0}^{\prime}}\sum_{\{\sigma_{i}\}=\pm1}\mathrm{e}^{\sum_{i}\ln(g_{r}(\sigma_{m,i}))-\beta\mathcal{H}_{m}^{\prime}(\{\sigma\})},\label{eq:Z_part}
\end{align}
 where $\mathcal{H}_{m}^{\prime}(\{\sigma\})$ is the Hamiltonian
of the effective spin-1/2 Ising model with non-symmetric degeneracy
given by eq.\eqref{hamt-s}.

For the purpose of mapping onto an exactly solvable model, we prefer
to change the eq.\eqref{eq:Z_part} onto an usual standard form through
a further transformation. Therefore, the partition function \eqref{eq:Z_part}
becomes

\begin{equation}
\mathcal{Z}_{m}(\beta)=\mathrm{e}^{\beta\mathcal{E}_{0}^{\prime}}\sum_{\{\sigma_{i}\}=\pm1}\exp(-\beta\sum_{<i,j>}\mathcal{H}_{i,j}^{\prime}),
\end{equation}
where 

\begin{align}
\mathcal{H}_{i,j}^{\prime}= & J\sigma_{m,i}\sigma_{m,j}-\frac{h}{\gamma}(\sigma_{m,i}+\sigma_{m,j})\nonumber \\
 & -\frac{1}{\gamma\beta}\left(\ln(g_{r}(\sigma_{m,i}))+\ln(g_{r}(\sigma_{m,j}))\right).\label{eq:H-p-sgm}
\end{align}

A further transformation, could leads us to an effective Ising model
Hamiltonian with temperature dependent field.

\subsection{The Ising model}

Finally using an additional transformation, the eq. \eqref{eq:H-p-sgm}
will be transformed onto a standard spin-1/2 Ising model, given simply
by 

\begin{equation}
\mathcal{\widetilde{H}}_{i,j}=\widetilde{J}\tau_{i}\tau_{j}-\frac{\widetilde{h}_{r}}{\gamma}(\tau_{i}+\tau_{j})+\tilde{E}_{r},\label{eq:H-t-sgm}
\end{equation}
at this stage, $\tau$ represents a standard spin-1/2, with $\widetilde{J}$,
$\widetilde{h}_{r}$ and $\tilde{E}_{r}$ being parameters to be determined.
Assuming the eqs.\eqref{eq:H-p-sgm} and \eqref{eq:H-t-sgm} are equivalents,
we have the following algebraic equations

\begin{align}
J-\frac{2h}{\gamma}-\frac{2}{\beta\gamma}\ln(g_{r}(1)) & =\widetilde{J}-\frac{2\widetilde{h}_{r}}{\gamma}+\tilde{E}_{r},\\
J+\frac{2h}{\gamma}-\frac{2}{\beta\gamma}\ln(g_{r}(-1)) & =\widetilde{J}+\frac{2\widetilde{h}_{r}}{\gamma}+\tilde{E}_{r},\\
-J-\frac{1}{\beta\gamma}\left(\ln(g_{r}(1))+\ln(g_{r}(-1))\right) & =-\widetilde{J}+\tilde{E}_{r}.
\end{align}

Solving this algebraic system equations, we obtain the following relation

\begin{align}
\widetilde{J} & =J,\\
\widetilde{h}_{r} & =h-\frac{1}{2\beta}\ln\left[\frac{g_{r}(-1)}{g_{r}(1)}\right],\label{eq:ht-fx}\\
\tilde{E}_{r} & =-\frac{1}{\beta\gamma}\ln\left[g_{r}(1)g_{r}(-1)\right].\label{eq:cnst-E}
\end{align}

It is worth to highlight that when mapping is symmetric, we have the
following relation $\tilde{h}_{r}=h$, similar to that discussed in
reference \cite{lett-exact}. 

In addition the partition function of the Hamiltonian \eqref{eq:Ham-sp-S},
is expressed by

\begin{equation}
\mathcal{Z}_{m}(\beta)=\mathrm{e}^{-\beta H_{0,r}}\sum_{\{\sigma_{i}\}=\pm1}\exp[-\beta(\widetilde{J}\sum_{\langle i,j\rangle}\tau_{i}\tau_{j}-\sum_{i}\widetilde{h}_{r}\tau_{i})],\label{eq:Z-Ising-h}
\end{equation}
where

\begin{align}
H_{0,r} & =-\mathcal{E}_{0}^{\prime}+\tilde{E}_{r}M\nonumber \\
 & =Nh\alpha_{0,m}-MJ\alpha_{0,m}^{2}-\tfrac{M}{\beta\gamma}\ln\left[g_{r}(1)g_{r}(-1)\right].
\end{align}

Some characteristic property of this model will be discussed now.
At high temperatures, $k_{B}T>2h/\ln\left(\frac{g_{r}(-1)}{g_{r}(1)}\right)$,
the term representing the effective field $\widetilde{h}_{r}=h-\frac{1}{2\beta}\ln\left[\frac{g_{r}(-1)}{g_{r}(1)}\right]$
is positive and thus the spins have a positive expectation values
$\langle\tau_{i}\rangle>0$. Whereas at low temperatures, $k_{B}T<2h/\ln\left(\frac{g_{r}(-1)}{g_{r}(1)}\right)$,
we have negative expectation values $\langle\tau_{i}\rangle<0$.

Miyashita et al.\cite{miyashita} discussed also the equivalence between
spin-crossover phase transition and that the metastable structure
of a charge transfer phase transition, where static metastability
exist in a study of the charge transfer transition in the material
$\mathrm{(nC_{3}H_{7})_{4}N[Fe^{II}Fe^{III}(dto)_{3}]}$ (dto=$\mathrm{C_{2}O_{2}S_{2}}$).

\section{The exactly solvable model}

In order to that eq.\eqref{eq:Z-Ising-h} becomes an exactly solvable
model, we consider the two-dimensional Ising model\cite{onsager},
so, this model can be solved exactly when $\tilde{h}_{r}=0$, therefore
it is equivalent to fix the magnetic field $h=\frac{1}{2\beta}\ln\left[\frac{g_{r}(-1)}{g_{r}(1)}\right]$
in eq. \eqref{hamt-s}. It is worth to note that, the magnetic field
$h$ only depends of the degeneracy of spin momenta and is proportional
to the temperature.

The free energy of non-bilinear spin-$S$ Ising model in thermodynamic
limit ($N\rightarrow\infty$ and $M=\gamma N/2$), can be written
in terms of standard spin-1/2 Ising model by the following relation

\begin{equation}
f_{m,S}=-\tfrac{1}{2\beta}\ln\left[g_{r}(-1)^{1-\alpha_{0,m}}g_{r}(1)^{1+\alpha_{0,m}}\right]+f_{1/2},
\end{equation}
where $f_{1/2}$ means the free energy of spin-1/2 Ising model.

Therefore, for two-dimensional case, we can consider three types of
lattice: triangular\cite{D&G V1}, square\cite{onsager} and honeycomb\cite{D&G V1}
Ising model.

The critical point $(\tilde{J}^{*},\tilde{h}_{r}^{*})=(\tilde{J}/T_{c},\tilde{h}_{r}/T_{c})$
for two-dimensional Ising model, in units of critical temperature
$T_{c}$, are given by 

\begin{equation}
\tanh(\tilde{J}^{*})=\begin{cases}
1-\sqrt{3}; & \text{triangular},\\
\sqrt{2}-1; & \text{square},\\
1/\sqrt{3}; & \text{honeycomb},
\end{cases}
\end{equation}
and $\tilde{h}_{r}^{*}=0$, for triangular, square and honeycomb lattice,
respectively.

The critical points for honeycomb ($\gamma=3$), square ($\gamma=4$),
and triangle ($\gamma=6$) lattice can be fully recovered, which is
consistent with the results previously obtained by Mi and Yang\cite{MiYang},
for the case of spin-1 Ising model (for detail see table I of reference
\cite{MiYang}). However our result is quite general and is valid
for any spin-$S$ and for any coordination number. 

The non-bilinear spin-$S$ Ising model critical points should satisfy
the relation below

\begin{alignat}{1}
B_{k}^{*}= & \left(\tfrac{1}{2}\ln\left(\tfrac{g_{r}(-1)}{g_{r}(1)}\right)-\gamma J^{*}\alpha_{0,m}\right)\alpha_{k,m},\label{eq:crit-Bs}\\
K_{k_{1},k_{2}}^{*}= & J^{*}\alpha_{k_{1},m}\alpha_{k_{2},m},\label{eq:crit-Ks}
\end{alignat}
by $^{*}$ we mean the parameters are in units of critical temperature
$T_{c}$. 

On the other hand, the term $\tilde{E}_{r}$ and $\mathcal{E}'_{0}$
are responsible for the appearance of residual entropy, in other words
this means due to competing parameters coupling of the non-bilinear
spin-$S$ model could be considered as a frustrated spin model, therefore
the entropy is given by

\begin{equation}
\mathcal{S}_{m}=\tfrac{1}{2}\ln\left[g_{r}(-1)^{1-\alpha_{0,m}}g_{r}(1)^{1+\alpha_{0,m}}\right].\label{eq:res-entr}
\end{equation}

Using the above result \eqref{eq:res-entr}, we can obtain a residual
entropy for the spin-1 Ising model discussed by Mi and Yang\cite{MiYang},
the first model has residual entropy given by $\mathcal{S}=\ln(2)$
while the second model has no residual entropy. It is worth to notice
the frustration properties of those model was not discussed by Mi
and Yang\cite{MiYang}.

Another simple example that we consider is the spin-3/2 Ising model,
for the particular case such that satisfy the eqs.(\ref{eq:sol321}-\ref{eq:sol325}).
The residual entropy for eqs. \eqref{eq:sol321} and \eqref{eq:sol322}
are zero, inasmuch as the independent coefficients of those polynomials
are $\alpha_{0,1}=\alpha_{0,2}=0$. However, for the last three polynomials
we have a residual entropy given by $\mathcal{S}=\ln(2)$, $\mathcal{S}=\frac{17}{16}\ln(3)$
and $\mathcal{S}=\frac{7}{16}\ln(3)$ for eqs. (\ref{eq:4}-\ref{eq:sol325})
respectively. It is interesting to highlight that, the residual entropy
is independent of the coordination number or some other lattice structure
parameters.

\section{Conclusion}

Different to those other methods developed to obtain this kind of
results using a more involved approach, we have used a simple spin-$S$
polynomial projection onto spin-crossover state, with non-symmetric
degeneracy of spin up or high-spin (HS) and spin down or low-spin
(LS). In general the present projection obtained have not be necessarily
symmetric with relation to their spins up or down. Only as particular
case of our results, we have the symmetric mapping which was previously
considered in reference \cite{lett-exact}, some additional results
are found also using the decoration transformation method satisfying
the 8-vertex model in our recent paper\cite{JPA-11}. Therefore we
conclude that, there is a spin-$S$ polynomial transformation onto
spin-crossover state, whose total possible number of projection is
given by $2(2^{2S}-1)$. Through a further transformation we can map
also onto a standard spin-1/2 Ising model.

As an application of this mapping we consider the non-bilinear spin-$S$
Ising model which can be transformed onto spin-crossover state described
by Wanjflasz model\cite{wanjflasz}. Using a further transformation
we obtain the partition function of the effective frustrated spin-1/2
Ising model\cite{onsager}, making a suitable mapping this non-symmetric
contribution leads us to a spin-1/2 model with a fixed external magnetic
field temperature dependent given by eq. \eqref{eq:ht-fx}. Therefore
we conclude that, the non-bilinear spin-$S$ Ising model such that
satisfy the projection proposed, must become equivalent to the Wanjflasz
model\cite{wanjflasz}, with quite interesting properties such as
residual entropy of the model, independent of the lattice structure.

\section*{acknowledgments}

O. Rojas and S.M. de Souza thanks CNPq and FAPEMIG for partial support.

\end{document}